\documentclass[namedreferences]{solarphysics}
\usepackage[optionalrh,solaromanenum,linksfromyear,natbib]{spr-sola-addons} 
\usepackage{graphicx}                    
\usepackage{color}                       

\newcommand{\degr}{\ensuremath{^\circ}}

\newcommand{\aap}{    {\it Astron. Astrophys.}}

\newcommand{\apj}{    {\it Astrophys. J.}}

\newcommand{\apjl}{   {\it Astrophys. J. Lett.}}

\newcommand{\grl}{    {\it Geophys. Res. Lett.}}

\newcommand{\solphys}{{\it Solar Phys.}}

\newcommand{\ssr} {	{\it Space Sci. Rev.}}

\begin{document}

\begin{article}

\begin{opening}

\title{Automatic Detection of Magnetic $\delta$ in Sunspot Groups}

%

\author{Sreejith~\surname{Padinhatteeri}$^{1,2}$\sep
        Paul A.~\surname{Higgins}$^{1,3}$\sep
	        D. Shaun~\surname{Bloomfield}$^{1}$\sep
		        Peter T.~\surname{Gallagher}$^{1}$
			       }

%
\runningauthor{Padinhatteeri {\it et al.}}
\runningtitle{Automatic $\delta$-spot detection}

%

\institute{$^{1}$ School of Physics, Trinity College Dublin, Dublin 2, Ireland 
			email: {sreejith.p@gmail.com} \\
		$^{2}$ 	Manipal Centre for Natural Sciences, Manipal University, Manipal, Karnataka, India - 576104. \\
	  $^{3}$ Lockheed Martin Advanced Technology Center, Dept/A021S,
     	     B/252, 3251 Hanover Street, Palo Alto, CA 94304, USA. }

\begin{abstract}

	Large and magnetically complex sunspot groups are known to be associated
	with flares. To date, the Mount Wilson scheme has been used to classify
	sunspot groups based on their morphological and magnetic properties. The
	most flare prolific class, the $\delta$ sunspot-group, is characterised by opposite
	polarity umbrae within a common penumbra, separated by less than 2\degr.
	In this article, we present a new system, called the Solar Monitor Active Region
	Tracker - Delta Finder (SMART-DF), that can be used to automatically
	detect and classify magnetic $\delta$s in near-realtime.  Using continuum
	images and magnetograms from the {\it Helioseismic and Magnetic Imager (HMI)}
	onboard NASA's {\it Solar Dynamics Observatory (SDO)}, we first estimate distances between opposite
	polarity umbrae. Opposite polarity pairs having
	distances of less that 2\degr\ are then identified, and if these pairs
	are found to share a common penumbra, they are identified as a magnetic
	$\delta$ configuration.  The algorithm was compared to manual $\delta$
	detections reported by the Space Weather Prediction Center (SWPC),
	operated by the National Oceanic and Atmospheric Administration (NOAA).
	SMART-DF detected 21 out of 23 active regions (ARs) that were marked as
	$\delta$ spots by NOAA during 2011\,--\,2012 (within $\pm$60\degr\
	longitude). SMART-DF in addition detected five ARs which were not announced
	as $\delta$ spots by NOAA. The near-relatime operation of SMART-DF
	resulted in many $\delta$s being identified in advance of NOAA's daily
	notification.  SMART-DF will be integrated with SolarMonitor\footnote{www.solarmonitor.org} and the
	near-realtime information will be available to the public.

\end{abstract}
%

\end{opening}

\section{Introduction}
     \label{S-intro}

Solar flares are among the most energetic events in the solar system
\cite[energy up to $\sim$$10^{25}$~J;][]{2012ApJ...759...71E,2001ApJ...552..833M, 2012ApJ...754..112A} influencing a panorama
of physical systems, from the solar surface through the heliosphere and onwards
into geo‐space.  Flares, along with coronal mass ejections (CMEs), are a major
contributor to space weather -- the interaction of magnetic fields and particles
accelerated on or near the Sun with the Earth's magnetosphere and upper
atmosphere \citep{2005GeoRL..32.3S01G, 2009SSRv..147..121M, ral2010}. Although
significant progress has been made in understanding the fundamental physics of
solar flares and coronal mass ejections, accurate forecasting of these enigmatic
events remains elusive \citep{2002SoPh..209..171G, 2006SoPh..237...45C,
2007ApJ...656.1173L, 2010RAA....10..785Y}.\\

Flares occur in the volume of atmospheric plasma above sunspot groups. Sunspot
groups are formed by the convective action of sub-surface fluid motions pushing
magnetic flux tubes through the Sun's surface, the photosphere. Turbulent
photospheric and sub‐photospheric motions jostle these flux tubes and, when the
conditions are right, the sunspots produce a flare \cite[ $e.g.$,][]{2008SoPh..248..297C}.
Currently the exact conditions that lead to flaring are not known.  The analysis
of sunspot groups and their properties has allowed the most accurate flare
prediction to date \citep{2008ApJ...688L.107B, 2012ApJ...747L..41B,
2013SoPh..283..157A}. \\

Studying large volumes of data to identify sunspot groups and their various
properties is generally done using feature-characterisation and tracking
algorithms.  The automatic detection of sunspots and their constituent structures
has been investigated using photospheric intensity images
\citep{zharkov2004,curto2008}  and magnetograms \citep{2005SoPh..228...55M,
2007ApJ...671..955L,2004SoPh..219...25L}.  Various feature tracking algorithms
were developed as part of the Heliophysics Events Knowledgebase
\citep[HEK;][]{hek}.  \citet{verbeek2013} provide a review of the robustness of
four major algorithms used to automatically detect and characterise active
regions (ARs) and sunspot groups.  The algorithms that they discuss are: the
Solar Monitor Active Region Tracker \citep[SMART;][]{2011AdSpR..47.2105H}, an
automated AR detection and characterisation algorithm that uses magnetograms to
detect magnetic features; the Automated Solar Activity Prediction code
\citep[ASAP;][]{2008SoPh..248..277C,2009SpWea...7.6001C}, a set of algorithms
that uses intensity images and machine learning to detect and predict flares;
the Sunspot Tracking and Recognition Algorithm
\cite[STARA;][]{2009SoPh..260....5W}; and the Spatial Possibilistic Clustering
Algorithm \citep[SPOCA;][]{2009A&A...505..361B}. A comparative study of flare
prediction between SMART-ASAP (a combination of the feature detection of SMART
and the machine learning of ASAP) and ASAP was performed in
\cite{2013SoPh..283..157A}, which found SMART-ASAP to be the more accurate.\\

Sunspot groups have traditionally been classified using the McIntosh
\citep{1990SoPh..125..251M} and Mount Wilson \citep{hale19} classification
schemes.  The McIntosh scheme classifies the complexity of sunspot groups from
their white-light structure, whereas the Mount Wilson scheme uses magnetograms
to classify the spatial distribution of magnetic polarities.

The original Mount Wilson classification scheme consisted of unipolar
($\alpha$), bipolar ($\beta$) and mixed polarity ($\gamma$) designations, before
being extended into the current Mount Wilson scheme through the addition of the
$\delta$ designation -- opposite polarity umbrae being enclosed by a common
penumbra \citep{1960AN....285..271K, 1965AN....288..177K}.  A statistical
investigation of the connection between $\delta$ spots and  major solar flares
was published by \cite{2000ApJ...540..583S}, finding that the vast majority of
X-class flares occur from sunspot groups that display a $\delta$ configuration
at some time.  However, Mount Wilson classifications are only issued every 24
hours by the National Oceanic and Atmospheric Administration (NOAA) Space Weather Prediction Center (SWPC), so it remains unclear
how long it takes to produce a major flare after a $\delta$ spot forms.
Automated detection of $\delta$ spot formation from high-cadence data is
necessary to answer this, with real-time implementation being beneficial for
flare forecasting. Here, a new software module named SMART-Delta Finder
(SMART-DF) is developed as an extension of the SMART algorithm to automatically
detect and characterise $\delta$ spots using simultaneous intensity and
magnetogram data.

In this paper,  the SMART-DF algorithm is explained in
Section~\ref{S-algo}. The SMART-DF code is tested on several individual cases, with Section~\ref{S-anal} presenting the
observation and data analysis.
Section~\ref{S-results} includes the results and
Section~\ref{S-conclusion} presents the conclusions and discussion of this work.

\begin{figure}
\centerline{\includegraphics[width=1.0\textwidth]{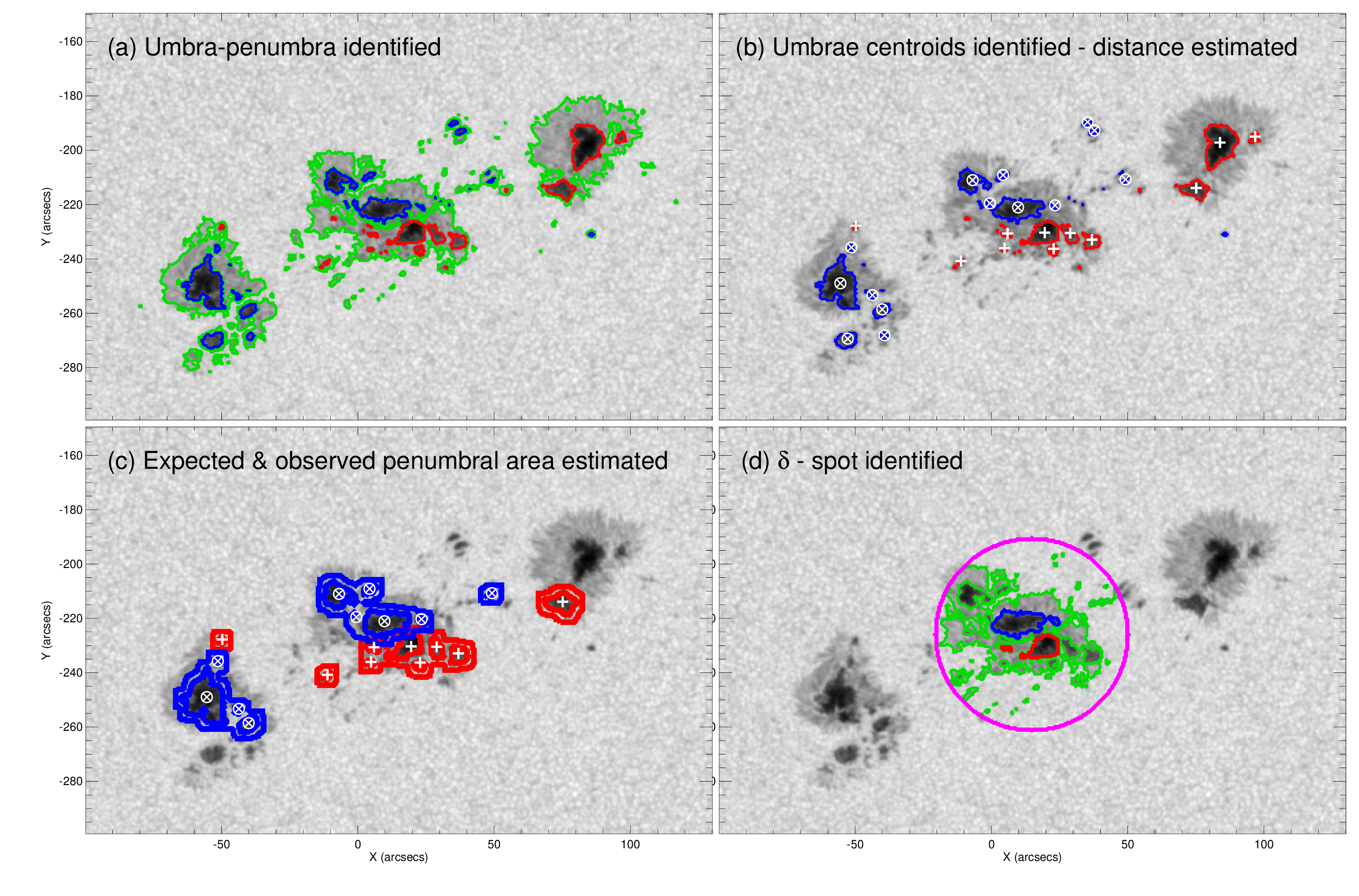}}
\caption{
Summary of the SMART-DF algorithm. (a) Umbra-penumbra boundaries are identified using
continuum intensities and LOS magnetic field values. Blue (Red) contours mark the negative
(positive) umbral border and green contours mark the penumbral boundary. (b) Each
umbra is labeled based on area and the centroids are identified and marked. 
A plus-symbol is used to denote positive and an encircled cross is used to
denote negative polarities.  Also
the distance between each positive-negative pair is estimated. (c) The boundary of
expected  penumbral regions around those positive (negative) umbrae which pass condition 1 (opposite polarity umbrae with in 2\degr) are marked with a red
(blue) contour. (d) If all conditions in the algorithm are satisfied, the identified
$\delta$ spot is marked with a magenta circle. See Section~\ref{S-algo} for a detailed
explanation of the SMART-DF algorithm.
} \label{fig-image1}

\end{figure}

\section{SMART-DF Algorithm}
 \label{S-algo}

SMART-DF is a software package developed using Interactive Data Language (IDL)
and SolarSoft \citep{solarsoft}.  The definition of the $\delta$ classification
as per \cite{1965AN....288..177K} is simply, a sunspot group with a ``common
penumbra enclosing opposite polarity umbrae''. Mt. Wilson Observatory uses a
more specific definition for $\delta$ spots as ``umbrae separated by less than
2\degr\  within one penumbra have opposite
polarity''\footnote{\url{www.swpc.noaa.gov/sites/default/files/images/u2/Glossary.pdf}}.
The only addition in this definition, compared to \cite{1965AN....288..177K}, is
the specific value for the maximum distance between two opposite polarity
umbrae. SMART-DF is developed based on this definition, and hence the distance
threshold between the two opposite polarity umbrae is kept as 2\degr.  This
algorithm implements the following two conditions for $\delta$ detection.
First, the region should have two opposite polarity umbrae with centroids
separated by less than 2\degr\ heliographic distance. Second, these two umbrae
should be surrounded by a common penumbra. These two conditions must be
satisfied to classify a sunspot group as a $\delta$ spot. In SMART-DF the
candidates that satisfy the first condition are named $\delta$ candidates, while
those that satisfy both conditions are labelled as $\delta$ spots.

A simultaneously observed photospheric intensity image and line-of-sight (LOS)
magnetogram are the necessary input to the code. The code also uses some
parameters which must be provided as input.  These include choosing a particular
field of view (FOV), instead of full disc, to detect $\delta$ spots.  When a
$\delta$ spot is found, SMART-DF will mark the location on an image, along with
an output structure providing details of magnetic and geometric properties of
the identified features. The values of some input parameters are critical, since
they decide the success of detection. The parameters and their optimised values
used in this study are listed in Table~\ref{tbl1}, and the reason for choosing
these values for each parameter are explained in Section~\ref{S-anal}.
Figure~\ref{fig-image1} shows the major steps of the SMART-DF algorithm, which
are described in the sections below.

\begin{figure}
\centerline{\includegraphics[width=0.9\textwidth]{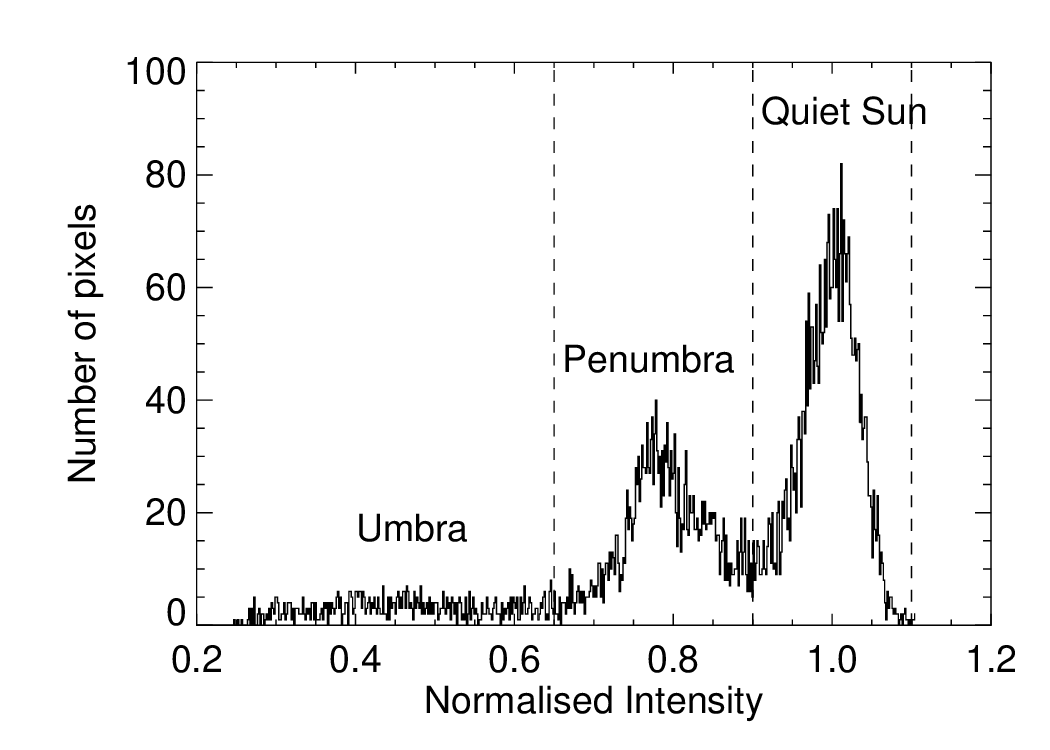}}
\caption{
A histogram of the intensity distribution of a mature sunspot that is normalised to the
quiet-Sun intensity. The peak at 1.0 corresponds to quiet-Sun pixels, and the
other two peaks correspond to penumbrae and umbrae, as indicated in the figure.
} \label{fig-cnthist}
\end{figure}

\subsection{Umbra-Penumbra Identification}

Regions of umbrae and penumbrae are identified using simultaneously observed
continuum and magnetogram images.  First, the intensity images are corrected for
limb-darkening using the \citet{1976asqu.book.....A} model. For each
intensity image, a simultaneous magnetogram is also obtained and
cosine-correction is applied to the magnetogram \citep{2005SoPh..228...55M}. A histogram of continuum intensity
for a region with a mature sunspot generally shows a triple-peaked distribution
as shown in Figure~\ref{fig-cnthist}. The Mean quiet-Sun intensity is defined as
$I_{\mathrm{QS}}$, and intensities less than 0.65\,$I_{\mathrm{QS}}$ correspond
to umbral pixels. Similarly, the peak at $\approx$0.77\,$I_{\mathrm{QS}}$
corresponds to the penumbral intensity \citep{leka98}. In this study, all pixels
with intensity less than 0.65\,$I_{\mathrm{QS}}$ and with a line-of-sight
magnetic-field intensity, $\left|B_{\mathrm{LOS}}\right|\geq$\,500\,G are
defined as umbral pixels.  Similarly, pixels within the intensity range of
0.65\,--\,0.9\,$I_{\mathrm{QS}}$ are defined as penumbral pixels and above
0.9\,$I_{\mathrm{QS}}$ are defined as quiet Sun. These values are used globally for all images. Also, since our results are not
dependent on flux within the umbral area, the errors in the intensity levels
used are not important. The above selection procedure is verified manually by
plotting  contours around the umbra and penumbra. An example can be seen in the
top-right image of Figure~\ref{fig-image1}.

\subsection{Distance Estimation}

To avoid detecting very small pores and inter-granular lanes, only umbrae above
a certain minimum cut-off area $({A}_{\mathrm{min}})$ are used to detect
$\delta$ spots. The value used for $A_{\mathrm{min}}$ in this study is given in
Table~\ref{tbl1}. Each umbra with an area larger than $A_{\mathrm{min}}$ in the FOV
is labeled based on its size and polarity. Hence, the largest positive-polarity
umbra will be labeled $U^+_1$, the next largest as $U^+_2, U^+_3,...,U^+_{n}$
and similarly, $U^-_1, U^-_2,...,U^-_{m}$ for the negative polarity. The
distance between each possible pairing of opposite polarity umbrae
 is calculated. The distance is measured from the center of the umbra, taken as the
$B_{\mathrm{LOS}}$ weighted centroid. Those pairs of umbrae with a distance of less
than 2\degr\ in heliographic coordinates are marked as first-level
$\delta$ candidates, which satisfy the first condition of the definition at the
beginning of this section. Note that each candidate is a pair of positive and
negative polarity umbrae, and only these candidates are considered for further
analysis.

\subsection{Presence of Common Penumbra around Spots}

Penumbrae form around the umbrae, with or without azimuthal symmetry. A region
around an umbral border with a certain width (the penumbral expected length,
$P_{\mathrm{el}}$) is selected and marked as the expected penumbral region
($P_{\mathrm{expected}}$). Actual penumbral filaments can be bigger or smaller
in size than $P_{\mathrm{el}}$ in different cases. Hence,
$P_{\mathrm{expected}}$ is simply the area around umbrae where we expect to find
penumbrae. Based on observed intensity levels, each pixel in this region is
marked as either penumbral or non-penumbral. The area of such observed penumbra
($P_{\mathrm{observed}}$) inside $P_{\mathrm{expected}}$ is calculated. This is
done for both the pair of umbrae in each $\delta$ candidate. Ideally, if
penumbrae are formed with azimuthal symmetry and fill the
$P_{\mathrm{expected}}$ region, then the ratio $P_{\mathrm{frac}} =
P_{\mathrm{expected}}/P_{\mathrm{observed}}$ should be equal to unity.  This is
not always the case.  In most of the cases the penumbra formation is not
azimuthally symmetric and in some cases the penumbral size can be less than the
width of the $P_{\mathrm{expected}}$ region, given by the parameter
$P_{\mathrm{el}}$. Hence, expecting such a ratio to be equal to unity is not
always practical. The ratio also depends on the value of $P_{\mathrm{el}}$ that
we choose as input. In this study we have used different values of this ratio to
qualify for second-level $\delta$ candidates, as explained in
Section~\ref{S-results}.

 The third level checks whether the penumbra around the pair of umbrae are
 connected. This is done by determining whether $P_{\mathrm{observed}}$ around
 each umbra in the pair is connected (sharing some common area,
 $P_{\mathrm{share}}$). The pairs of opposite polarity spots that pass the three
 checks are marked as $\delta$ spots. This algorithm becomes unreliable when the
 active region approaches the limb, due to projection effects of both
 foreshortening and apparent ({\it i.e.,} false) LOS magnetic-field polarity
 inversions. In this paper, we test SMART-DF only for ARs within $\pm$60\degr\ 
 longitude.

\section{Observations and Data Analysis}
 \label{S-anal}

SMART-DF was tested using data from the {\it Helioseismic and Magnetic Imager}
\citep[HMI;][]{hmi2012main}. The HMI instrument was developed by the Stanford Solar
Group and is a part of NASA's {\it Solar Dynamic Observatory} (SDO). HMI can be used
to obtain continuum intensity images as well as LOS magnetic-field strength of
the full solar-disc at high temporal cadence (every 45 seconds). HMI also
obtains magnetic-field vector information using polarimetry, but this is not used
in this study. The primary input to SMART-DF is a pair of simultaneous continuum
intensity images and magnetograms.  All the ARs that  appear on the solar-disc
during 2011 and 2012, with in $\pm$60\degr\ longitude were analysed to test the
reliability of SMART-DF. A pair of intensity  images and LOS magnetograms were
obtained with 12 hour cadence, i.e., at 00:00 UT and 12:00 UT every day from the
archived HMI data.  SMART \citep{2011AdSpR..47.2105H} is used to automatically
identify ARs and define regions of interest (ROIs) on the solar-disc. All ARs located within $\pm$60\degr\ of longitude have been checked for
$\delta$ configurations using SMART-DF as explained in Section~\ref{S-algo}.
To calculate distances and areas on the solar-disc, firstly helioprojective coordinates are  converted to heliographic coordinates using the world coordinate-system \citep[WCS;][]{wcsref} programs \citep{2006A&A...449..791T} available in the standard solar data analysis software (SolarSoft SSWIDL) and then spherical-trigonometry cosine-law is used \citep{smartBook}.

SWPC, operated by NOAA, releases a Solar Region Summary (SRS) text
file every day at 00:30 UT. SRS files contain the Hale classification of each AR
on disc. NOAA classifies each active region by visually analysing the previous
day data. Each day SRS between 1 January 2011 and 31 December 2012 were obtained from
the SWPC database and a list of ARs that formed $\delta$ configurations was created.
Detections of $\delta$ spots from SMART-DF were compared with this list and a
comparison test was performed using two methods.

The first method tests whether all ARs detected by NOAA as having a $\delta$
configuration  on a particular day were also detected by SMART-DF. For this,
those days on which a new AR appeared on the solar-disc were noted down (based
on NOAA numbers in the SRS files). Each AR that formed a $\delta$ configuration
according to the SRS files during 2011 and 2012 were listed and the date of its first
detection as a $\delta$ was noted. SRS is based on observations of the Sun the
previous day to the `release date'. Hence, SMART-DF was used to check for
$\delta$ configurations in the data of that day (the previous day of the SRS
release date that has a $\delta$ spot). In this test, SMART-DF is limited to
find less than or an equal number of $\delta$ spots compared to SRS. The results
have been compared and are presented in Section~\ref{S-results} of this paper.

The second method tests the relative detection rates of NOAA/SWPC and
SMART-DF. On each day, the number of $\delta$ configurations identified by
NOAA are counted and plotted as a histogram distribution. Full-disc images
(simultaneous continuum and magnetogram) are obtained with a 12-hour cadence,
meaning twice per day for each day in 2011 and 2012. SMART-DF was used to look
for $\delta$ configurations and a similar histogram was created and over-plotted
for comparison. SMART-DF can detect less than, equal to or more instances of
$\delta$ configurations NOAA/SWPC detections.

\begin{table}
\begin{center}
\caption{The parameters used by SMART-DF and their optimised values used for
this algorithm testing study.}
\begin{tabular}{|c|c|l|}
    \cline{1-3}
 Parameter& Value Used& Details\\
    \cline{1-3}
    $A_{\mathrm{min}}$ & 1.5 Mm$^2$ & Minimum/cut-off umbral area \\
    $P_{\mathrm{el}}$ & 3.5 Mm & Expected length of penumbra/rudimentary penumbra \\
    $P_{\mathrm{frac}}$ & 75\% & Ratio of actual penumbra with expected penumbral area\\
    $P_{\mathrm{share}}$ & 750 km$^2$ & Common shared area between penumbrae of two spots\\
    \cline{1-3}

\end{tabular}
\label{tbl1}
\end{center}
\end{table}

There are four important input parameters needed to obtain reliable results from
SMART-DF, as explained in the previous section (Section~\ref{S-algo}). They are:
\begin{enumerate}
    \item $A_{\mathrm{min}}$, the minimum umbral area to be
	counted as an umbra;
    \item $P_{\mathrm{el}}$, the expected length of a
	penumbra (or rudimentary penumbra, as it is called in the case of a
	forming penumbra);
    \item $P_{\mathrm{frac}}$, the fraction of penumbral
	area observed with respect to the expected penumbral area.
    \item $P_{\mathrm{share}}$, the common area shared by the penumbrae of two opposite
	polarity spots.
\end{enumerate}

	The optimisation of these four parameters is critical for successful
	$\delta$ spot detection. The optimised values used for this study are
	given in Table~\ref{tbl1}.  The optimisation was performed by varying these parameters to find the combinations with the  maximum successful detection rate of NOAA detected regions. Users are free to change these values in the code to suit their needs or if they find any anomalous AR going undetected ({\it e.g.}, see the discussion on NOAA AR 12192 in Section~\ref{S-conclusion}). The minimum umbral area required to qualify as an umbra is fixed at 1.5 Mm$^2$, which is greater than the typical size-scale of granulation \cite[known to be $\sim$1 Mm
	;][]{2009lrsp2}.  The expected length of the penumbra (or rudimentary
	penumbra) is fixed at 3.5\,Mm, the typical size of a forming penumbra
	\citep{solanki2003}. Technically the value of $P_{\mathrm{el}}$ could be as large as the minimum distance that is needed to pass condition 1 of a $\delta$  spot ({\it i.e.}, 2\degr\ in heliographic coordinate system).  However, intergranular lanes and other intensity regions seen in complex active regions can be wrongly marked as penumbra  when $P_{\mathrm{el}}$ is too large. Hence, it is important to choose an optimal value that matches with typical penumbral length-scales. The third parameter is the fraction of actual
	penumbral area with respect to expected penumbral area. Ideally this has
	to be equal to unity ({\it i.e.}, 100\%). However, the actual penumbral length
	can be different from case to case, and in many cases penumbrae will not
	be formed completely around both umbrae. Hence if we fix this value to
	100\%, even if there is a small region without penumbra, the algorithm
	will fail to count them as a $\delta$ candidate. This fraction can be
	significantly different from case to case, so SMART-DF was  tested with
	different values of $P_{\mathrm{frac}}$ on a known $\delta$ region. This
	resulted in an optimal value of 75\% for $P_{\mathrm{frac}}$ which was
	used in the rest of this study. This value also depends on the value of $P_{\mathrm{el}}$ chosen. The fourth parameter,
	$P_{\mathrm{share}}$, is fixed at 750\,km$^2$ (2 pixels $\times$ 2
	pixels in the observed data). This is the common area shared between two
	penumbrae, so that they appear as one common penumbra.

\begin{figure}
\centerline{\includegraphics[width=1.0\textwidth,clip=]{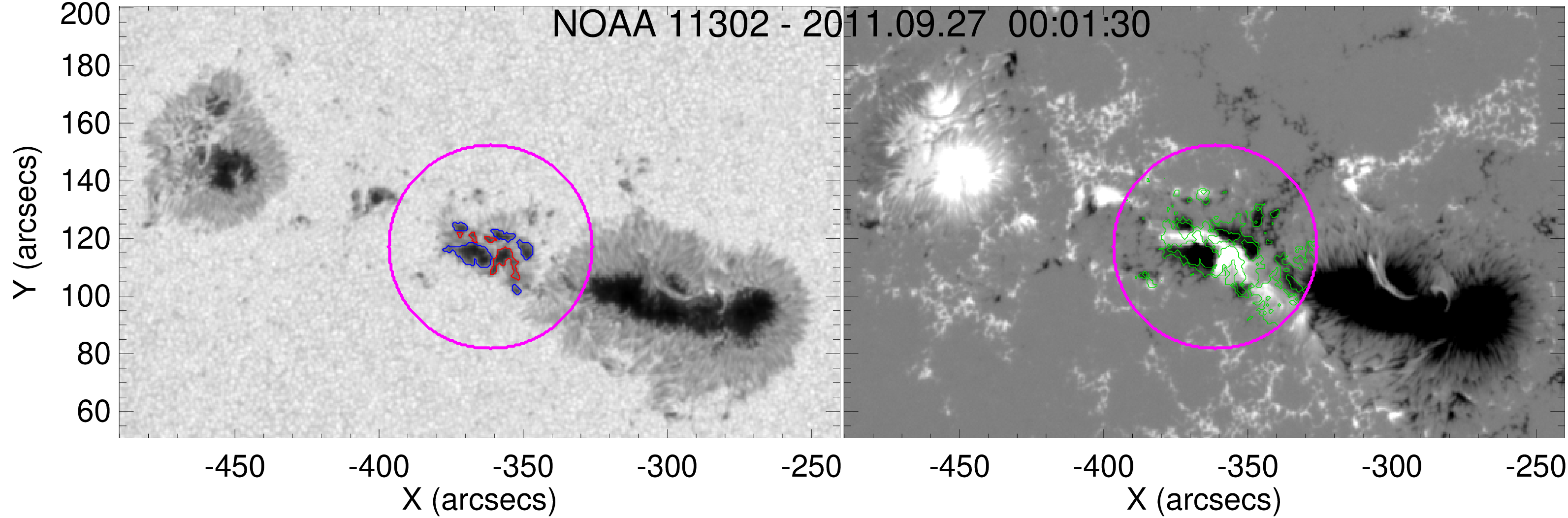}}
\centerline{\includegraphics[width=1.0\textwidth,clip=]{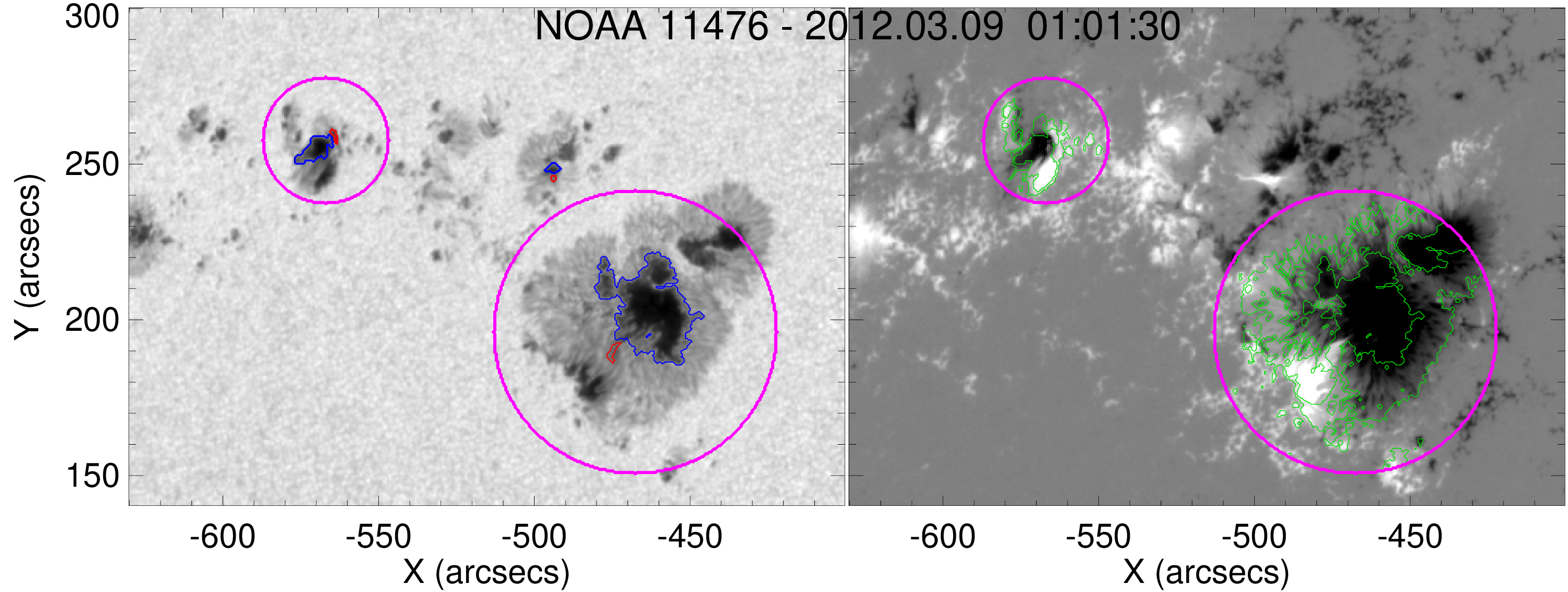}}
\centerline{\includegraphics[width=1.0\textwidth,clip=]{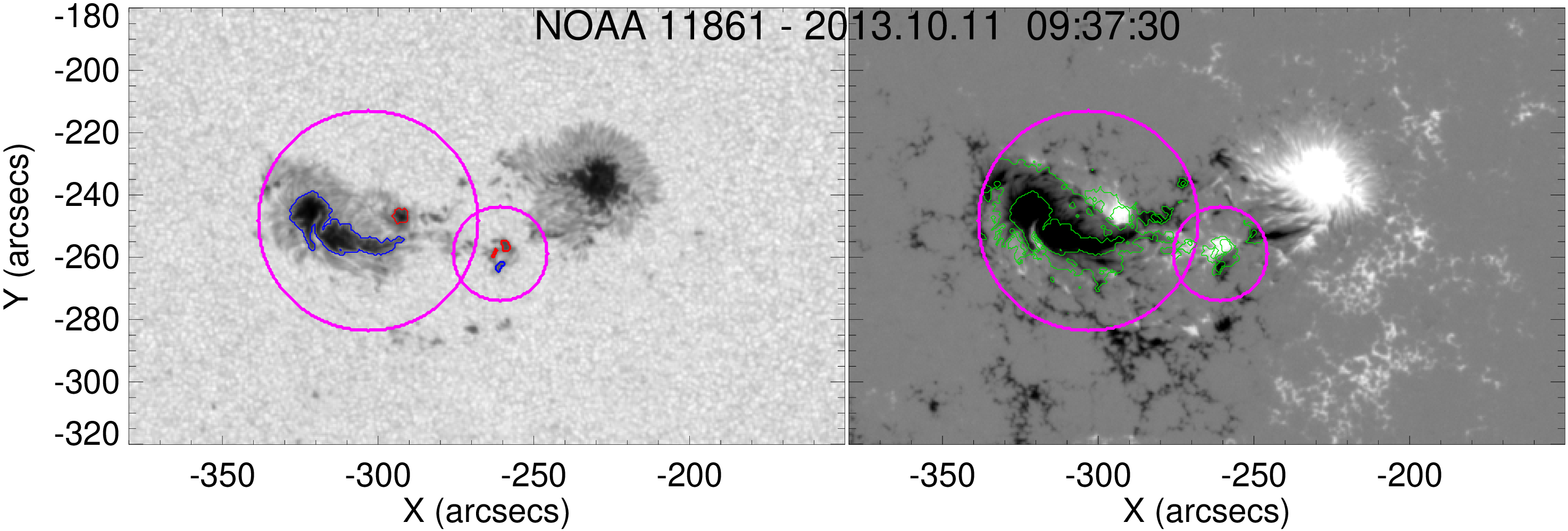}}
\caption{An example of a $\delta$ spot detected using the automatic code. The red (blue)
contours show the positive (negative) umbrae and the green contours mark the
penumbral border. A pink circle surrounds the location of a detected $\delta$ configuration.
\label{fig-result1}}
\end{figure}

\section{Results }
 \label{S-results}

The SMART-DF algorithm was tested on multiple ARs. One example can be seen in
the bottom-right image in Figure~\ref{fig-image1}. Three other examples of
$\delta$ spot detection are shown in Figure~\ref{fig-result1}. In each row, the
left side shows an HMI photospheric intensity image and the right side shows an 
HMI LOS magnetogram.  The positive (north) and negative (south) polarity umbral
areas are marked with red and blue contours, respectively. Green contours mark
penumbral areas. A circle is drawn to highlight the region of interest centred
at the $\delta$ forming region.

In the upper row in Figure~\ref{fig-result1}, two main spots with opposite
polarity are observed, with the leading spot having negative polarity and the
following spot having positive polarity. The $\delta$ formation occurs in the
region of flux emergence between the two main spots. It is also observed in this
case that the $\delta$ spot is formed by multiple small umbrae with opposite
polarity and having a rudimentary penumbra around the whole region. A similar
configuration is observed in the emerging region, but with two distinct, large
and stable bipolar spots in the bottom right image of Figure~\ref{fig-image1}.  The
central row of the Figure~\ref{fig-result1} shows another case of two distinct
$\delta$ spot formations in the same AR. One occurs in the emerging-flux region,
similar to the earlier case. Another $\delta$ is formed by a small
positive-polarity spot joining with a large negative leading-spot. Both spots
share a common penumbra surrounding both polarities.  The fourth example of a
$\delta$ spot detected by SMART-DF is given in the bottom row of
Figure~\ref{fig-result1}. Two $\delta$ forming pairs can be seen here. One of
them is similar to the above example, where a small positive-polarity spot joins with a large, stable negative-polarity spot. In some cases opposite polarities of a small bipolar region may approach each other and form a rudimentary penumbra for a short time. SMART-DF
indicates them as a $\delta$ since they satisfy both required conditions, and
hence are marked with a small circle in the figure (bottom row).
Such cases are generally seen at locations of flux emergence. A user can choose
to avoid the detection of such small-scale and transient $\delta$ spots by
increasing the cut-off umbral-area threshold ($A_{\mathrm{min}}$).

The success rate of $\delta$ detection was also tested by comparing with the
detections by NOAA/SWPC. There are two methods of performing this comparison, as
explained in Section~\ref{S-anal}. The first method of testing is to check
whether each AR is detected by both NOAA and SMART-DF on its first day of
observation.  During 2011\,--\,2012, NOAA's SRS files reported 33 ARs as having
a $\delta$ pair at some part of their appearance on the solar-disc. Out of
these, 23 ARs formed a $\delta$ within $\pm$60\degr\ longitude. SMART-DF
detected 21 out of these 23 NOAA detections several hours before the SRS was released by
NOAA. The algorithm failed to detect any  $\delta$ regions in NOAA AR 11164 and NOAA AR 11374 when compared to SRS. Both were manually checked, on AR 11164 there was not enough penumbra at the beginning when the opposite polarities were within 2\degr, and by the time the penumbra developed the opposite polarities drifted away to more than the cut-off distance of 2\degr. In the case of AR 11374 the negative polarity around the positive spot was spread like a plage and did not form a sunspot with area more than $A_{\mathrm{min}}$.

\begin{figure}
\centerline{\includegraphics[width=0.9\textwidth]{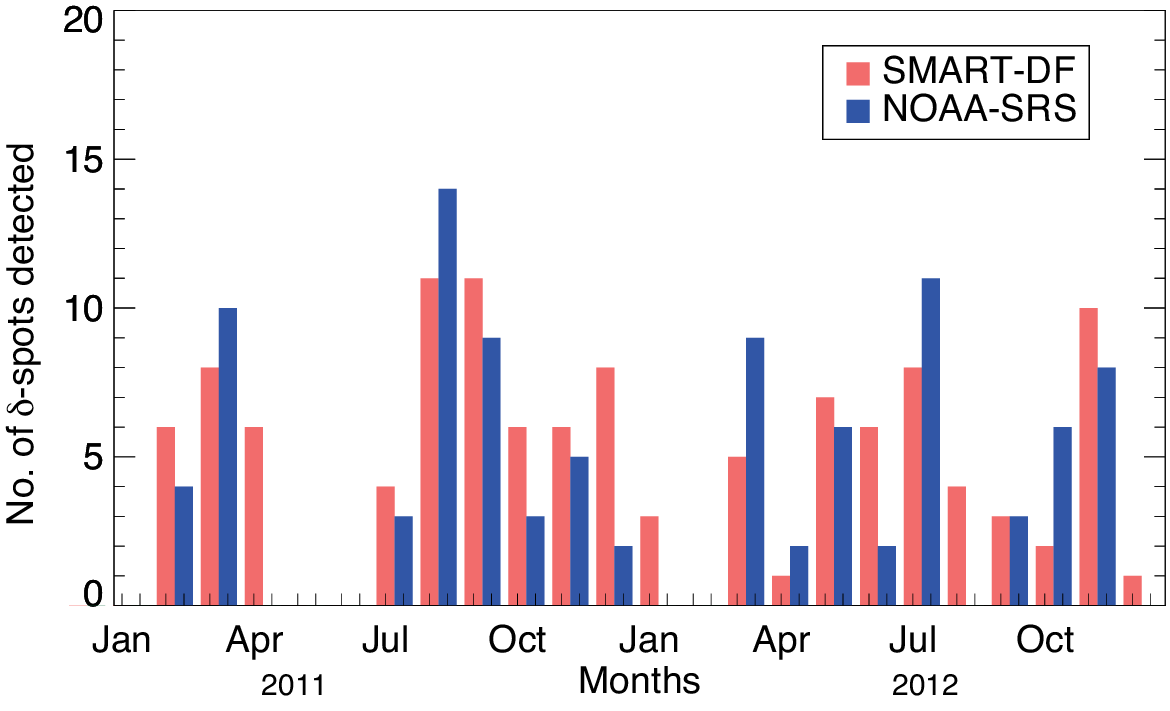}}
\caption{
Number of $\delta$ spots detected by SMART-DF and NOAA every month of 2011\,--\,2012. Red bars represent the SMART-DF detection and blue bars
represent the detections by NOAA SRS. SMART-DF detected 116 instances of
$\delta$ spots when compared to 97 by NOAA during 2011\,--\,2012.
} \label{fig-result2}

\end{figure}

The second method of testing is to compare the number of instances of
daily $\delta$ detections by NOAA and SMART-DF. The
$\delta$ configuration in an AR may remain so for more than one day,
and it will be counted again as another instance by both SMART-DF and NOAA. To
match with NOAA's daily SRS, SMART-DF counts each $\delta$ region only once {\it per}
day (though SMART-DF uses a cadence of two observation {\it per} day).
During the two years studied here, NOAA detected 97
instances of $\delta$ formation within $\pm$60\degr\ of longitude while
the SMART-DF detected 116 cases. Distributions of the two are given in
Figure~\ref{fig-result2}. In these additional detections by SMART-DF, five ARs
were never reported by NOAA. It is possible that NOAA ground-based data did not show a
$\delta$ spot, due to its inferior spatial resolution or possible increased
seeing. Since only two instances {\it per} day (at 00:00 UT and 12:00 UT)
were used by SMART-DF for analysis, if a $\delta$ spot formed and disappeared
between the two instances (with a gap of 12 hrs), SMART-DF would have missed
that case, while NOAA would have detected them as they scan through out the day.
A higher cadence data analysis would show a better match in the distribution. At
the same time, SMART-DF may detect small, instantaneous $\delta$s which may be
missed, or ignored by manual analysis by NOAA. When multiple
$\delta$ configurations are formed in the same AR, both SMART-DF and NOAA list count
them as one.

\section{Discussion and Conclusions}
     \label{S-conclusion}

	In this article we report on the development of SMART-DF, an algorithm to
	automatically detect $\delta$ spots in active regions. These forms of
	sunspot groups are known to be associated with major solar flares
	\citep{2000ApJ...540..583S}.  Current methods to identify $\delta$ spots
	are primarily based on manual checking of ground-based data ({\it e.g.},
	NOAA).  Our near-realtime detection of the formation of $\delta$ spots
	can be used to flag an AR as a potential flaring region. SMART-DF will
	be integrated into SolarMonitor\footnote{www.solarmonitor.org}, and hence will be available to the public.

     The results of comparing SMART-DF and NOAA/SWPC $\delta$ spot detection
     rates are presented. It is found that during 2011 and 2012, 21 out of 23
     ARs detected by NOAA as $\delta$ spots (within $\pm$\,60\degr\ longitude)
     were identified by SMART-DF during the 24-hour period before the SRS release by
     NOAA (which occurs at 00:30 UT every day). In some cases SMART-DF detected
     $\delta$ spots an entire day (during 24\,--\,48 hours period) before it was
     marked as a $\delta$ by NOAA.  In addition SMART-DF detected five ARs which
     were never reported by NOAA as $\delta$ region. If daily instances of
     $\delta$ detections were to be counted, 97 instances were detected by NOAA
     compared to SMART-DF 116 cases. SMART-DF might have detected more
     $\delta$ spots if the data had been analysed using higher cadence. In this
     study, data were obtained only once in 12 hrs, any instance of short term
     $\delta$ formation and disappearance during the interval are missed.


\begin{figure}
\centerline{\includegraphics[width=1.0\textwidth,clip=]{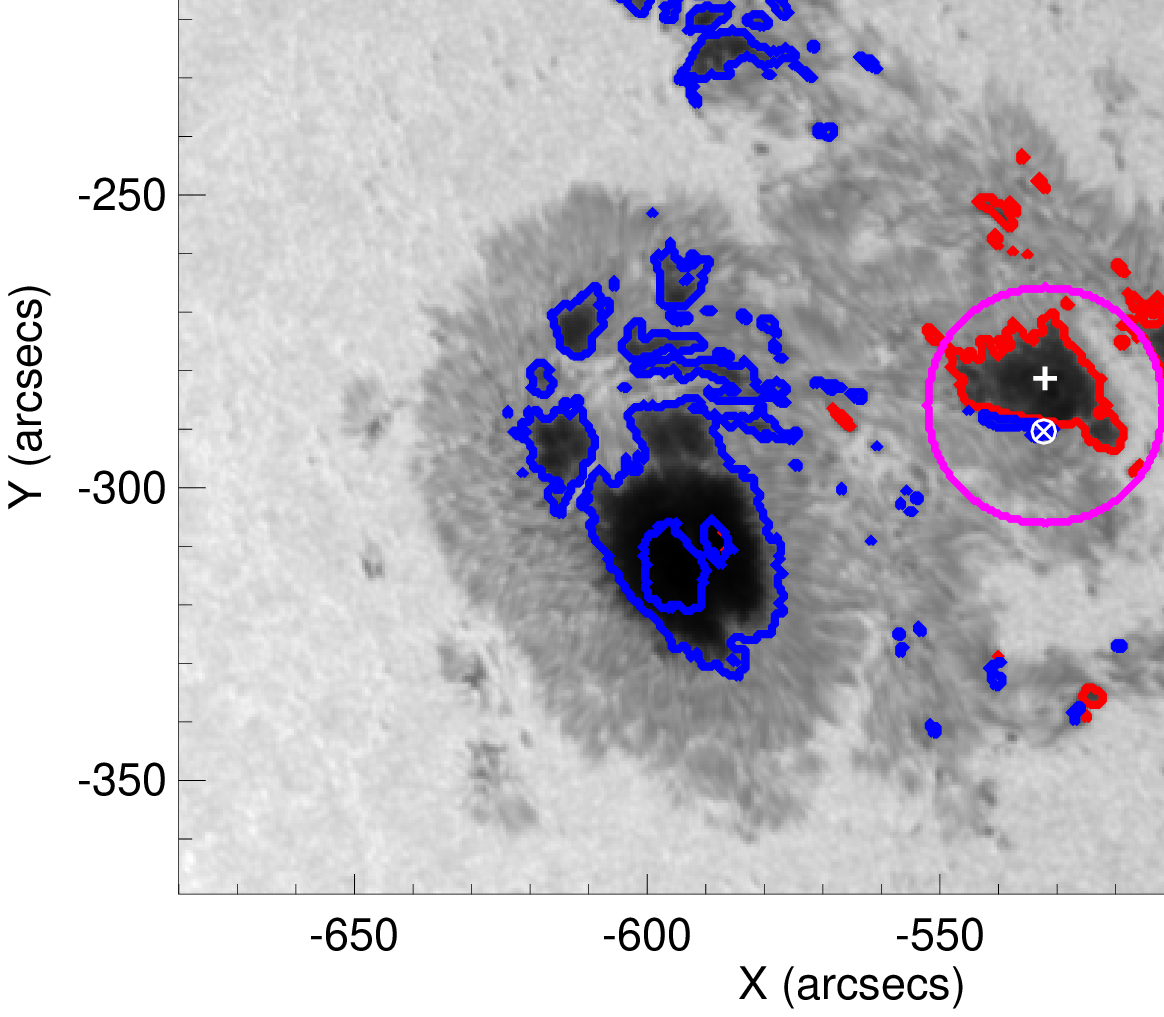}}
\centerline{\includegraphics[width=1.0\textwidth,clip=]{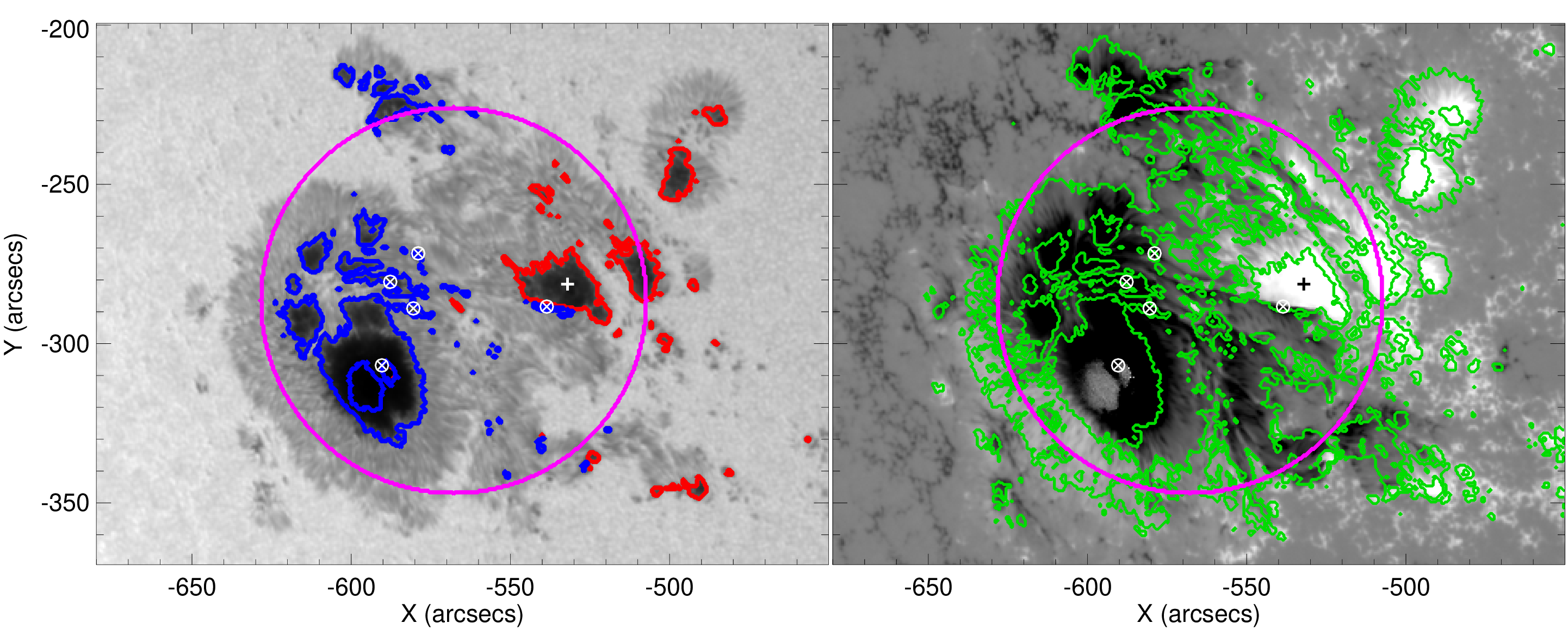}}
\caption{ HMI observation of AR12192 and it's analysis by SMART-DF . The top and
bottom rows of the figure represent two cases with different values for the
maximum allowed distance between opposite polarity umbrae (to be detected as a
$\delta$ spot). The top row is with standard definition value of 2\degr, and the
bottom row is with a value of 5.2\degr. Red (blue) contours show the positive (negative)
umbrae and green contours mark the
penumbral border. A circle surrounds the location of a detected $\delta$ configuration.
\label{fig-12192}}
\end{figure}

    SMART-DF determines that a spot is a $\delta$ configuration when two
    conditions are met: i) opposite polarity umbrae within 2\degr and ii) common
    penumbra surrounding both polarity umbrae. However, the conventional
    definition of $\delta$ spots is vague. While the Mt. Wilson
    Observatory uses the specification of within 2\degr\ separation, the initial
    definition by \cite{1960AN....285..271K, 1965AN....288..177K} does not
    mention any value.  The recent appearance of the extremely large region NOAA AR 
    12192 was one specific case where the distance between opposite polarity
    umbrae (both between their centroids and between their closest edges) is
    more than 2\degr. As shown in the top row of Figure~\ref{fig-12192},
    SMART-DF fails to find the correct $\delta$ spot pair ({\it i.e.}, the two largest umbrae), instead selects a dark and very small negative-polarity region on
    the edge of the positive-polarity umbra. The correct pair of umbrae was
    detected only when the restriction of distance was increased, with the
    bottom row of Figure~\ref{fig-12192} the result when separations of up to
    5.1\degr\ are allowed. Although large ARs like AR 12192 are not common, there
    may be several candidates in which SMART-DF failed because the umbrae
    separation was slightly more than 2\degr. In addition, the definition of
    penumbrae has changed significantly during the last decade with the advent
    of high-spatial resolution observations. In earlier observations, any
    shadow-like region around an umbra (with intensity less than the quiet Sun, but
    more than the umbra) was called penumbra. More recently, penumbrae are known to
    have distinct filamentary structure with a well defined magnetic topology.
    SMART-DF, like most other algorithms, uses only intensity values around
    umbrae to identify penumbrae.  When two pores are in close proximity, a
    shadow-like region can form between them with an intensity between that of
    the quiet Sun and the pores ({\it i.e.}, umbrae).  Currently, SMART-DF will
    identify this as a penumbra and may use that to satisfy the second
    $\delta$ spot condition.

    The detection of a $\delta$ spot does not assure the occurrence of a flare,
    but for flare forecasting studies it will be important to determine how useful
    $\delta$ spot detection is in identifying potentially flaring ARs.
    In a follow-on study, SMART-DF will be used to investigate the evolution of different physical
    parameters in $\delta$ spots, including the
    observed changes in magnetic topology and plasma velocity.

%

%

%
 \begin{acks}

This work has received financial support from: EOARD (SP), Irish Research
Council- Enterprise partnership (PAH),
European Space Agency Prodex programme (DSB).
Courtesy of NASA/SDO and the HMI science teams for the data used in this paper.

 \end{acks}

%
 \bibliographystyle{spr-mp-sola}
%

\end{article}
\end{document}